\documentclass[landscape]{aa}
\usepackage{graphicx,psfrag,amsmath,lscape}

\newcommand{\kms}{$\mathrm{\,km\,s}^{-1}$ }

\newcommand{\Vt}{$\mathrm{V_{t}}$ }
\newcommand{\Teff}{${T_{\rm eff}}$ }
\newcommand{\logg}{$\log \mathrm{g}$ }
\newcommand{\FeH}{$\mathrm{[Fe/H]}$ }
\begin{document}
\title{A new B\"{o}hm-Vitense gap in the temperature range 5560 to 5610
  K in the Main Sequence
\thanks{Based on spectra collected with the ELODIE spectrograph at the
1.93-m telescope of the Observatoire de Haute Provence (France).}
\thanks{Full Table 1 is only available in electronic form at
http://www.edpsciences.org}
}
\titlerunning{New B\"{o}hm-Vitense gap in the Main Sequence}
\author{V.V. Kovtyukh \inst{1},\, C. Soubiran \inst{2},\, and S.I.
 Belik \inst{1}}
\authorrunning{Kovtyukh et al.}
\offprints{V.V. Kovtyukh, \\
           e-mail: val@deneb.odessa.ua}
\institute{
Astronomical Observatory of Odessa
National University and Isaac Newton Institute of Chile,
Shevchenko Park, 65014, Odessa, Ukraine
\and
Observatoire Aquitain des Sciences de l'Univers, CNRS UMR 5804, BP 89,
33270 Floirac, France
 }
\date{Received 10 June 2004 / Accepted 20 July 2004}
\abstract{
Highly precise temperatures ($\sigma = 10-15$ K) have been determined from
line depth ratios for a set of 248 F-K field dwarfs
of about solar metallicity (--0.5 $<$ [Fe/H] $<$ +0.4),
based on high resolution
($R$=42\,000), high $S/N$ echelle spectra.
A new gap has been discovered in the distribution
of stars on the Main Sequence in the temperature range 5560 to 5610 K.
This gap coincides with a jump in the microturbulent velocity \Vt
and the well-known Li depression near 5600 K in field dwarfs
and open clusters. As the principal cause of the observed
discontinuities in stellar properties
we propose the penetration of the convective zone
into the inner layers of stars slightly less massive than the Sun
and related to it, a change in the temperature gradient.
 }

\maketitle

\section{Introduction}
Of the fundamental stellar parameters, effective temperature (\Teff)
is the most crucial to obtain accurate abundances. There
exists a multitude of methods to determine \Teff, among which we
have chosen the line depth ratios. This method has the advantage of being
independant of interstellar reddening, spectral
resolution and rotational and microturbulence broadening. In a
previous paper (Kovtyukh et al. \cite{kovtet03}) we presented high
precision temperatures for 181 F-K dwarfs by this method. We have
noticed a gap in the temperature range 5560 to 5610 K
($B-V\sim$0.70 for about solar metallicity)
that we investigate in this paper.

Color gaps are known to exist in open clusters.
For example, high-precision astrometry and photometry of the Hyades
reveal two gaps
along the clusters Main Sequence (MS), near $B-V\sim$0.38
and $\sim$0.48 mag
(de Bruijne, Hoogerwerf \& de Zeeuw \cite{dBHZ00}; \cite{dBHZ01}).
Rachford \& Canterna (\cite{rachcan00})
describe them in several other clusters. Among field stars, the existence
of such gaps is still matter of debate (e.g.,
B\"{o}hm-Vitense \& Canterna \cite{bohm74}; Mazzei \& Pigatto \cite{MP88};
Simon \& Landsman \cite{SL97}).
Newberg \& Yanny (\cite{NY98}) have used $Hipparcos$ high-precision
parallaxes,
$V$ and $B-V$, to construct the H-R diagram of nearby stars (primarily
field stars, although their sample may contain cluster stars).
They only found evidence for a gap at $B-V\sim$0.25 in the MS.

The explanations for gaps range
from the onset of convection in the stellar envelope
(B\"{o}hm-Vitense \& Canterna 1974) to peculiarities in the  Balmer jump
and Balmer lines as suggested by Mermilliod (\cite{mer76}).
Ulrich (\cite{ulrich71a}; \cite{ulrich71b})
speculated that gaps in young clusters might be produced
by the presence of $^{3}$He isotopes that halt
gravitational contraction 1...2 mag above the MS. While Mermilliod's
gaps occur among B stars, B\"{o}hm-Vitense \& Canterna's
(and also Ulrich's) gaps are found among less massive stars.
Note that the standard stellar models do not show any peculiar change
of slope in the \Teff versus mass relation along the MS.
B\"{o}hm-Vitense (\cite{BV70}) was the first to realize that
a gap in the distribution of colors may develop when the atmosphere
changes from purely radiative to convective in the deep layers
at \Teff cooler of $\sim$ 7500 K.
There are two possible ways in which
convection might alter \Teff: (a) either because the convective temperature
gradient is steeper than radiative
(B\"{o}hm-Vitense  \cite{BV70}); or (b) because convection causes
photospheric granulation
with associated temperature inhomogeneities
(Nelson \cite{Nelson80}; B\"{o}hm-Vitense \cite{BV82}).

In order to investigate the existence of gap in the MS, we have
increased our initial sample to 248 stars, with \Teff ranging from 4000 to
6150 K and absolute magnitudes $M_{V}$ from 3.02 to 7.41.
In Sect. 2, we briefly describe our dataset and the method of line
depth ratios. In Sect. 3, we confirm the existence of a gap in the
temperature range 5560 to 5610 K and we show that it is not correlated
with [Fe/H] but with the microturbulence velocity \Vt.
Also we discuss the possible coincidence of this gap with
the Li depression observed in field stars and open clusters.

\section{Data reduction and the derivation of atmospheric parameters}

The spectra used in this paper were extracted from the most recent
version of the library
of stellar spectra collected with the ELODIE echelle spectrograph
at the Observatoire de Haute-Provence by Soubiran et al. (\cite{soubet98})
and Prugniel \& Soubiran (\cite{prusou01}).
The description of the
instrument mounted on the 1.93 m telescope can be found in
Baranne et al (\cite{baran96}).

Most of the stars have accurate $Hipparcos$ parallaxes enabling
as to determine their absolute magnitudes M$_{V}$ that range from
3.016 (\object{HD 209965}, F8V) to 7.410 (\object{HD 29697}, K3V).
\object{HD 209965} however has the
most uncertain absolute magnitude due to a relative error in parallax
of 18\% (all the others have relative errors lower than 11\%).
Two stars that are not part of the $Hipparcos$ catalogue had
their absolute magnitudes estimated by the TGMET method introduced by
Katz et al. (\cite{katzet98}): \object{HD 171304} (M$_{V}$=4.238),
\object{HD 186039} (M$_{V}$=4.026).

The spectral range of the spectra is
4400--6800 \AA\AA\, and the resolution is $R$=42\,000.
The signal to noise ratio ($S/N$) of the spectra range from 40 to 680, with
the large majority having $S/N>$ 100.
Spectrum extraction, wavelength
calibration  and radial velocity measurement were performed
at the telescope with the on-line data reduction software while
straightening of the orders, removing of cosmic ray hits, bad pixels and
telluric lines
were performed as described in Katz et al (\cite{katzet98}).
 Further processing of the spectra (continuum placement, measuring
equivalent widths, etc.) was carried out by us using the DECH20 software
(Galazutdinov \cite{gal92}). Equivalent widths and depths
$R_{\lambda}$  of lines were measured manually by means of a
Gaussian fitting. The Gaussian height was taken as a measure of the
line depth. This method produces line depth values that agree
with the parabola technique used by Gray (\cite{gray94}).

The key to our analysis of the stellar spectra
is the line-depth ratio technique,
pioneered by Gray (1994) and successively improved by our group.
This technique allows the determination of \Teff with an exceptional
precision. It relies on the ratio of the central
depths of two lines that have very different functional dependences
on \Teff. The method is independent of the
interstellar reddening and only marginally dependent on individual
characteristics of stars, such as rotation, microturbulence and metallicity.

Briefly, a set of 105 line ratio -- \Teff relations was obtained
in Kovtyukh et al. (\cite{kovtet03}), with the mean random error
of a single calibration being 60--70 K
(40--45 K in most cases and 90--95 K in the least accurate
cases). The use of $\sim$70--100 calibrations per spectrum reduces
the uncertainty to
5--7 K. These 105 relations use
92 lines, 45 with low ($\chi<$2.77 eV) and 47 with high ($\chi>$4.08 eV)
excitation potentials, and have been calibrated with the 45 reference
stars in common
with Alonso et al. (\cite{AAMR96}), Blackwell \& Lynas-Gray
(\cite{BLG98}) and di Benedetto (\cite{dB98}). The zero-point of
the temperature scale was directly adjusted to the Sun
(\Teff=5777 K is adopted for the Sun), based on  11 solar
reflection spectra taken with ELODIE, leading to an uncertainty
in the zero-point of about 1 K. The application range
for the calibrations is $-0.5<$\FeH$<+0.4$.

In the present study we add 82 new spectra to the 2003 sample,
increasing the total number of dwarfs with precisely determined
\Teff to 248 (see Table 1). In Table 1 we give the
mean \Teff, the number of calibrations used ($N$), the errors of the
mean $\sigma$, \logg, \Vt, [Fe/H], M$_{V}$ and $B-V$ from $Tycho$2
(transformed into Johnson).
For the majority of stars we get an error which is smaller than 10 K.

\begin{table*}
\caption[]{Program stars.}
\begin{tabular}{ccccccrccc}
\hline
 Star&\Teff , K & $N$& $\sigma$, K& \logg &\Vt, \kms & [Fe/H]& $M_{V}$ &(B--V)&Remarks \\
\hline
 HD 1562   & 5828&   97&  5.8 &  4.0 & 1.2 &--0.27&  5.007 &  0.589&                 \\
 HD 1835   & 5790&   68&  5.5 &  4.4 & 0.9 &  0.20&  4.843 &  0.644&                 \\
 HD 3765   & 5079&   87&  4.7 &  4.3 & 1.1 &  0.06&  6.155 &  0.953&                 \\
 HD 4307   & 5889&   91&  5.0 &  4.0 & 1.1 &--0.13&  3.637 &  0.583&                 \\
 HD 4614   & 5965&   69&  6.4 &  4.4 & 1.2 &--0.19&  4.591 &  0.542&                 \\
\hline
\end{tabular}
\end{table*}

Accurate [Fe/H], \logg and \Vt have been determined as well.
The iron abundance was determined from an LTE analysis of
equivalent widths of FeI and FeII lines
using the WIDTH9 code and the atmosphere models by Kurucz (\cite{kur93}).
Appropriate models for each star were derived by means of standard
interpolation through \Teff and \logg.

Microturbulent velocities \Vt were determined by forcing
abundances obtained from individual FeI lines to be independent
of equivalent width. The value of
\Vt is very sensitive to the adopted \Teff.
The small uncertainty of 0.1-0.2 \kms is the result
of our \Teff estimates being accurate to 10-20 K.

Surface gravities \logg were derived from the
ionization balance between FeI and FeII.

The abundance of iron relative to solar
\FeH=log(Fe/H)$_*$ -- log(Fe/H)$_\odot$
was adopted as the metallicity parameter of a star and was obtained
from FeI lines.
Note that the solar metallicity from our analysis is [Fe/H]=0.05
(Table 1). This non-zero value results from the lower resolution of our
spectra ($R$ = 42\,000) compared to the  resolution of the Solar Spectrum
Atlas ($R\sim$600\,000) which is usually used for solar oscillator strengths
and [Fe/H]$_{\odot}$ determinations.
For some stars the atmospheric parameters
have been previously reported in Mishenina et al. (\cite{MSKK04}).

\section{The gap}

Fig. 1 shows the resulting temperature versus absolute magnitude
for our 248 field dwarfs. A number of turn-offs
and a well-defined MS can be seen.
The latter shows a clear gap between \Teff=5560 and 5610 K
($B-V\sim$0.70 mag for about solar
metallicity) and $M_{V}$=5.0--5.5 mag,
which corresponds to a mass of about 0.90--0.95 M$_{\odot}$.
This gap has been tentatively identified in Hyades by
De Bruijne, Hoogerwerf \& de Zeeuw  (\cite{dBHZ00}, \cite{dBHZ01}),
although it is less prominent than the other two gaps
in this cluster (since Hyades has [Fe/H] = +0.15, this leads
to a shift of the $(B-V)_{gap}$ to 0.75--0.80).

The gap (near $(B-V)\sim$0.70) is not
clearly seen in the $(B-V)-M_{V}$ diagram for our sample of field stars,
because the $(B-V)$ parameter is very sensitive to interstellar
reddening, metallicity and macroturbulence.

What is the probability that the gap is due to statistical uncertainty
in the star counts ?
The histogram appears to be a smoothly increasing function
between \Teff=4900 and 5900 (Fig. 2a). Approximating by a
quadratic polynomial, we get
a dispersion in the distribution of about $\pm$3.2 (point (5585,0) included)
or $\pm$2.0 (point (5585,0) excluded). This predicts 12 stars
in the region of the gap (centered on \Teff=5585 K).
No star in the gap requires a minimum of 4$\sigma$ deviation, which
translates into the probability that the gap is real as high as 0.999.

Note that one star (excluded from Table 1) does fall into the gap.
\object{HD 13974} is a G0V spectroscopic binary,
with \Teff=5590 $\pm$12.9
and [Fe/H]=--0.46. The value of \logg=3.9 for this star is more typical
of a subgiant, as well as M$_{V}$=4.697.
We therefore conclude it is not a dwarf.
Also quite possibly the spectrum of the secondary component
distorts our measurements of the lines of the primary
component and therefore affects the derived \Teff.

\begin{figure}[hbtp]
\begin{center}
\includegraphics[width=7cm]{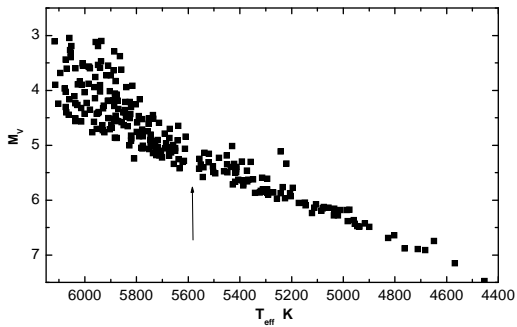}
\caption{Temperature-absolute magnitude diagram for 248 field dwarfs.
The errors in the temperature are less than the symbol size.
An arrow indicates the gap, which may be caused by the sudden
changes in the properties of atmospheric convection near \Teff=5560 K.
}
\label{f:ab_tg}
\end{center}
\end{figure}

\begin{figure}[hbtp]
\begin{center}
\includegraphics[width=7cm]{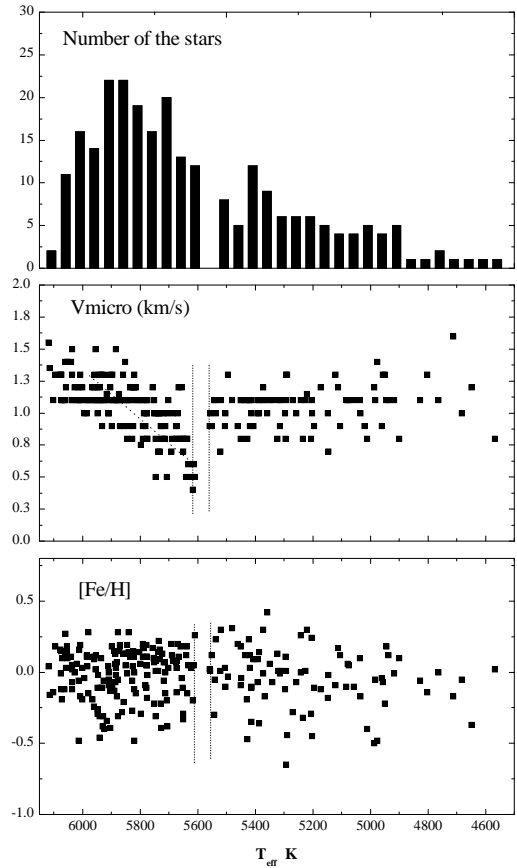}
\caption{(a)-- A histogram of field dwarf distribution with temperature.
The bin size is 50 K, starting from the minimum temperature of 4510 K;
(b)-- Microturbulent velocity versus \Teff;
(c)-- Iron abundance vs. \Teff.}
\label{f:ab_tg}
\end{center}
\end{figure}

It is reasonable to assume that microturbulence is driven
by convection and therefore one might expect a discontinuous behavior
of \Vt across the gap.
Indeed, as seen from Fig. 2b, \Vt experiences a jump at the
position of the gap. On the hot side, \Vt declines from 1.3 \kms
at \Teff=6000 K to 0.6 \kms at 5560 K. On the cool side, \Vt is
nearly constant at 1.0 \kms.
This test has not been carried out before and may be an indirect
confirmation of the convection hypothesis.
It is very important to note that this jump remains even if
we decrease T$_{eff}$ by 50 $K$ for the stars located near the hot boundary
of the gap. Therefore we conclude that this jump is real.

To demonstrate that there is no systematic effect in our \Teff
or \Vt determination, we plot the derived metallicities versus \Teff
in Fig. 2c. There is no trend of [Fe/H] with \Teff
which confirms the robustness of our atmospheric analysis.
Also, it shows that our sample is very uniform in metallicity,
[Fe/H] spanning the full interval at all temperatures.

The dwarfs in our sample have various ages, and they likely
evolve within the MS.
Why then the position of the gap does
remain independent of the stars' ages and their evolutionary stages?
One possible solution is that. The structure of the outer parts of the
convective zones is independent of
small luminosity changes during the  MS stage, but depends mainly upon
the \Teff. Therefore, stars in the vicinity of the gap should experience
the color and \Teff jumps, placing them outside the gap.
Thus, we can consider these jumps as a result of the sudden rebuilding of
the outer convective zone structure that change the color and \Teff of the
stars.

\subsection{The possible lithium -- gap connection}

The gap is located near the hot edge
of the well-studied lithium depression in the field dwarfs at \Teff$<$5600,
where lithium is depleted by 2 or more orders of magnitude
(see, for example, Chen et al. \cite{chen01}; Ryan \cite{Ryan00}).
The dwarfs located on the left of the blue side of the gap show a
large spread
in lithium abundance, with values A$_{Li}$ ranging from less than
1.0 to 3.0 dex. In contrast, stars located on the right of red side show
low values of A$_{Li}$, reflecting the dilution effects
in this spectral region.

The surface
abundance of Li should in principle be a function of \Teff, age,
metallicity and somewhat poorly explored physical processes like
convection, rotation, stellar wind, etc. Thus, by measuring Li
in stars with nearly solar metallicity, crucial information
about the convection in stars of different ages can be obtained.
The level of dilution of lithium depends on the level
of convection in these stars.

Li depression is observed in field stars and
old clusters such as Coma, \object{Hyades}, \object{M 67}
but not in the younger ones like \object{Pleiades} and Alpha Per.
The location of Li depression
is also age-dependent: in the Hyades (625 Myr) it occurs
at \Teff$\sim$5400 K
(Boesgaard \& King \cite{BK02}), in \object{NGC 752} (1.7 Gyr)
at \Teff$\sim$5600 K, and in M 67 (5 Gyr) at
\Teff$\sim$5700 K (Ryan 2000).
Only stars that have shallow convection regions will show Li enrichment -
when a star has a large convective region, the Li is diluted
beyond detection.
Cool dwarfs (\Teff $<$5600\,K) usually show lower Li abundance
compared to hot ones. Only a few cool dwarfs are known to have a high
Li content (Chen et al. 2001).

It is clear that the sudden decline in A$_{Li}$ near the
temperature gap is directly associated with a rapid increase in thickness
of the convective envelope in such regions
(e.g. Cayrel, Cayrel de Strobel \& Campbell \cite{cayrel84}; Chen et al. 2001).
Most of the stars with high
Li abundance present an undeveloped convective zone, whereas stars with
low Li abundance show a developed convective zone.

Israelian et al. (\cite{ISMR04}) find that lithium
abundances of planet host stars
with \Teff between 5850 and 6350 K are similar to those in the
Chen et al. (\cite{chen01}) comparison sample,
while at lower \Teff, in the range 5600 -- 5850 K, the planet host stars
show on average lower
lithium abundances than stars in the comparison sample.
This temperature range is where anomalous behavior of the
microturbulence velocity is observed (Fig. 2b).

\section{Summary}

To investigate the fine structure in the distribution
of stars on the MS,
we have determined precise \Teff for a sample of
248 field F,G and K stars
with accurate $Hipparcos$ parallaxes and near-solar metallicity
$-0.5<$[Fe/H]$<$0.4.
We have identified a new B\"{o}hm-Vitense
gap in the MS between 5560 and 5610 K, with a confidence
better than 99.9\%.
This gap possibly coincides with the Li depression  seen in open
clusters and field stars.
The most likely explanation for the gap is the sudden change
in the properties of atmospheric convection.

Our analysis suggests that the
theoretical models of convection in solar-type stars need to be improved.

\begin{acknowledgements}
V.K. wants to thank the staff of Observatoire de Bordeaux for the
kind hospitality during his stay there.
In addition, V.K. would like to thank Prof. G.S. Bisnovatyi-Kogan
and Dr. N.I. Gorlova for many useful discussions about stellar physics.
The authors are grateful to the referee, Dr. B.A. Twarog, for helpful
comments and suggestions which have certainly improved the final
version of this paper.
\end{acknowledgements}

\end{document}